# Googling the Big Lie: Search Engines, News Media, and the US 2020 Election Conspiracy

Ernesto de León[1], Mykola Makhortykh[2], Aleksandra Urman[3], and Roberto Ulloa[4]

[1]University of Amsterdam

[2]University of Bern

[3]University of Zurich

[4]GESIS - Leibniz Institute for Social Sciences

Corresponding author: Ernesto de León, e.deleon@uva.nl, orcid: 0000-0003-3152-0722

March 2024

**Googling the Big Lie: Search Engines, News Media, and the US 2020 Election Conspiracy**

**Abstract:** The conspiracy theory that the US 2020 presidential election was fraudulent – the Big Lie – remained a prominent part of the media agenda months after the election. Whether and how search engines prioritized news stories that sought to thoroughly debunk the claims, provide a simple negation, or support the conspiracy is crucial for understanding information exposure on the topic. We investigate how search engines provided news on this conspiracy by conducting a large-scale algorithm audit evaluating differences between three search engines (Google, DuckDuckGo, and Bing), across three locations (Ohio, California, and the UK), and using eleven search-queries. Results show that simply denying the conspiracy is the largest debunking strategy across all search engines. While Google has a strong mainstreaming effect on articles explicitly focused on the Big Lie – providing thorough debunks and alternative explanations – DuckDuckGo and Bing display, depending on the location, a large share of articles either supporting the conspiracy or failing to debunk it. Lastly, we find that niche ideologically driven search queries (e.g., 'sharpie marker ballots Arizona') do not lead to more conspiracy-supportive material. Instead, content supporting the conspiracy is largely a product of broader ideology-agnostic search queries (e.g., 'voter fraud 2020').

# 1 Introduction

To the date of writing, American ex-president Donald Trump continues to promote the conspiracy theory that the 2020 presidential election was stolen from him. A tale weaving together misinterpretation of events, exaggerations of minor incongruencies and fabricated allegations, it claims that Joe Biden won through voter fraud. The pervasiveness of this story led to it being awarded its own nickname – the

Big Lie. Despite being unable to provide evidence of these claims – Trump's lies have not only been thoroughly debunked, but the numerous lawsuits filed have all been struck down, and recounts in contest states have confirmed previous results – a large portion of the American population still believes that the election was fraudulent. Depending on the poll, between 60 to 80% of Republicans do not believe that Joe Biden was legitimately elected, almost two years after the election (Arceneaux & Truex, 2021; Fahey & Alarian, 2022). Furthermore, the strong link between the Big Lie and the January 6th Insurrection shows how a single conspiracy theory can strike at the heart of US democracy.

   The news media have a complicated relationship with conspiracy theories, especially the Big Lie. Being gatekeepers to information, the news media hold the power of both correcting misguided narratives of malicious plots and of oxygenating the narratives that feed into conspiracy beliefs. How news media communicate about conspiracies and engage with the claims they make is therefore crucial. This is especially the case for a conspiracy with the vocal support of prominent political leaders, such as Republican elites who have unapologetically promoted claims of voter fraud, using the conspiracy as a key campaign message, and writing legislation in its stead (Arceneaux & Truex, 2021). Furthermore, the numerous lawsuits surrounding the case and the large-scale investigation by the January 6th Select Committee put the Big Lie squarely on the media agenda. Journalists have to therefore choose how to engage with the unfounded narratives surrounding the conspiracy.

Much work has been dedicated to the strategies journalists use to counter conspiracies and misinformation (for a meta-analysis of studies, see Walter & Murphy, 2018), with distinctions drawn between the simple negation of statements versus full debunking and alternative explanations. Nevertheless, much less research has explored the role search engines play in exposure to these strategies. Today, search engines are a key part of information exposure on the web. They are a regularly used – and highly trusted (Pan et al., 2007; Schultheiß, Sunkler, & Lewandowski, 2018) – source of news access (Moller, Van De Velde, Merten, & Puschmann, 2020), with citizens turning to them directly in search of information. While some research has explored the role these algorithmic gatekeepers have in exposing users to conspiracies (Rovetta & Others, 2021; Urman, Makhortykh, Ulloa, & Kulshrestha, 2022), we know little about how they deal with a conspiracy that becomes a high-profile and regular part of the media agenda as the Big Lie. Similarly, while there is an emerging field studying how search algorithms filter news content (Moller et al., 2020; Ulloa & Kacperski, 2022), it has not focused on news covering conspiracies. Therefore there is a significant gap in our understanding of how search engines prioritize different types of news coverage on a conspiracy that is part and parcel of the US news cycle, and exposure to which has significant implications for the US democracy.

To address this gap, we examine the relationship between search engines and online exposure to conspiracy news as well as the role of news media in the debunking of conspiracy theories, using the Big Lie as a case study. Specifically, we explore whether the use of different search engines leads to encountering distinct sources of news, relevant articles on the topic, and debunking strategies. Because of concerns over location-based personalization of search results (Kliman-silver, Hannak, Lazer, Wilson, & Mislove, 2015), we compare how results vary across two different regions within the US and one region in the UK. Additionally, we engage with the recent work raising concerns over differences in search terms used by partisans (Trielli & Diakopoulos, 2022) which experts worry might lead to politically-consonant results (Tripodi, 2022; Van Hoof, Meppelink, Moeller, & Trilling, 2022).We do so by investigating the differences between search results in response to niche ideologically motivated search queries (e.g. 'sharpie marker ballots arizona') and more neutral terms (e.g. 'voter fraud 2020'). We do so by conducting a large-scale algorithmic audit of the news output content produced by searching for the US 2020 voter fraud conspiracy. For this purpose, a total of 360 virtual agents were deployed across three search engines (Google, DuckDuckGo, and Bing), in three different locations (Ohio, California, and the UK), using eleven different conspiracy-related search terms. A content analysis is then used to identify the relevance of the news item for the conspiracy, as well as the debunking strategies applied by journalists.

## 2    Theoretical Framework

### 2.1    The Big Lie, Conspiracies, and the News Media

Conspiracies usually lurk in the shadows of niche online communities, with research showing their strong presence on social media and blogs (Wood & Douglas, 2015). Nevertheless, the mainstream news media do play a role in spreading conspiracy theories. While research has shown that news media often portray conspiracy theories in a negative light (Uscinski & Parent, 2014), they still serve as the place of first contact for many individuals (Arnold, 2008; Stempel, Hargrove, & Stempel Iii, 2007; Stieger, Gumhalter, Tran, Voracek, & Swami, 2013). In fact, exposure to conspiracy theories in the media has been linked to higher receptivity to their ideas (Arnold, 2008; Dimaggio, 2022), especially if the conspiracy has ideological underpinnings (Vegetti & Mancosu, 2020; Warner & Neville-shepard, 2014). Compounding this effect is the fact that the Big Lie differs from most other conspiracies in that it has occupied more of the media agenda than other conspiracies. Its promotion by some members of the Republican elite, its clear links to the January 6, as well as the ongoing legal disputes that have spawned as result has meant that it is a regular feature in the news (Dimaggio, 2022; Painter & Fernandes, 2022). While journalists need to report on socially relevant events, they also have a responsibility to provide the public with correct information, helping minimize the spread of such misinformation. Studies have emphasized the importance of debunking and fact-checking when covering conspiracies (Cook, Ecker, & Lewandowsky, 2015). However, the effectiveness of these strategies varies considerably.

A common strategy for countering the spread of conspiracy theories is fact-checking. Here, we define fact-checking as preceding or following misleading statements with rebuttals, informing readers that a statement is incorrect, or that there is no evidence to back such claims. Common in political journalism, this strategy of explicitly verifying specific statements and/or narratives is employed to expose the falsehood of information behind a conspiracy to the reader (Ecker, Lewandowsky, & Tang, 2010). Nevertheless, studies have documented how readers can continue to remember false information even if it was fact-checked, in a process known as the continued influence effect (Cook et al., 2015).

Furthermore, the rise of polarization can render fact-checks ineffective due to motivated reasoning, a process exacerbated in political conspiracy theories (Druckman & Mcgrath, 2019). These fact checks also run the risk of engaging in potential 'both sides-ism' by trying to present a balanced picture, with the disputed conspiracy theory and the fact check being seen as two sides of a political debate (Bisgaard & Slothuus, 2018; Clarke, 2008).

Fact-checking strategies can also fall short of convincing their audience due to the promotion of uncertainty. While negating the veracity of a specific account, fact-checks can fail to provide the necessary detail and offer an alternative explanation for the piece of misinformation (Ecker, Lewandowsky, & Apai, 2011). In fact, such fact-checks have been shown to cause a 'backfire' effect (Nyhan & Reifler, 2010), where a reader's belief in misinformation is reaffirmed; however, more recent work has challenged this idea, especially in the political domain (Nyhan, 2021).

A second strategy lies in appeals to coherence. This strategy mainly relies on providing alternative explanations for misinformation and conspiracies (Cook et al., 2015): by informing readers not only that a statement is untruthful but also explaining what the reality is and the motivation behind actors' misinformation, journalists are more likely to convince readers in an enduring manner (Johnson-laird, Gawronski, & Strack, 2012). The psychology behind such effects relies on the cognitive need to organize facts logically (Gilovich, 2008) – a simple omission can threaten a mental model of unfolding events (Anderson, Lepper, & Ross, 1980; Johnson & Seifert, 1994), leaving a cognitive gap in understanding (Cook et al., 2015; Gerrie, Belcher, & Garry, 2006). By filling that gap with an alternative explanation, arguments become more coherent and are therefore more easily processed, believed, and retained (Walter & Murphy, 2018). These corrections have been documented to be even more effective when journalists document the motivation behind the conspiracy (Lewandowsky, Stritzke, Oberauer, & Morales, 2005). A meta-analysis of studies on corrections showed that this strategy was more effective at reducing misinformation and conspiracy beliefs (Walter & Murphy, 2018). The close scrutiny and negative coverage that accompanies appeals to coherence have also been shown to help elevate the costs of elite misinformation (Snyder Jr & Stromberg, 2010), therefore reducing the prevalence of such conspiracies by prominent politicians (Nyhan & Reifler, 2015).

Although not a debunking strategy, another way journalists can reduce the spread of conspiracy theories is by omitting any conspiracy-related information Journalists have the ability to draw lines between their stories and conspiracies, echoing elite misinformation on a seemingly unrelated topic. This is especially prevalent when conspiracies and misinformation are coming from political elites, as its usually coupled with partisan cues (Mccright & Dunlap, 2011). For example, stories about voter reform following the 2020 elections have been linked to the Big Lie (Arceneaux & Truex, 2021). However, journalists can also choose not to amplify misinformation that feeds into conspiracies, and are not crucial to a story. By omitting such statements, journalists can "avoid amplifying false claims and reduce the incidence of partisan and ideological cues when discussing matters of fact and science" (Nyhan, 2021, 4).

Lastly, some news outlets actively promote belief in conspiracies. Studies have shown that hyperpartisan and alternative news sites have been shown to contain stories promoting conspiracies (Davis, 2009), while others have shown that readership of these sites is linked to conspiracy beliefs (Stempel et al., 2007).

## 2.2    (Conspiracy) Information Exposure on Search Engines

Search engines serve as key information intermediaries in today's high-choice media environment. By filtering and ranking information in response to user queries, they determine which sources get more visibility and what information the users are exposed to. As users highly trust web search results (Pan et al., 2007; Schultheiß et al., 2018), search engines become agenda-setters, shaping perceptions of the broad range of subjects, from gender and race (Noble, 2018; Urman & Makhortykh, 2022) to elections (Trielli & Diakopoulos, 2022; Urman, Makhortykh, & Ulloa, 2021) to historical phenomena (Makhortykh, Urman, & Ulloa, 2021; Zavadski & Toepfl, 2019). At the same time, the quality of web search outputs varies widely across topics and search engines. To examine the quality of search results, scientists commonly rely on algorithm impact auditing - the process of systematically investigating outputs of complex algorithmic systems with the aim of evaluating them for the presence of systematic distortions (Mittelstadt, 2016). Previous studies have shown that the search outputs can be biased in terms of gender and race (Noble, 2018; Urman & Makhortykh, 2022) and that the information provided in top search results is not always factually correct, including in highly consequential domains such as health or politics (Makhortykh, Urman, & Ulloa, 2020; Urman et al., 2022).

The importance of search engines for the public sphere is amplified by them being important gateways for news. Using web tracking data, (Moller et al., 2020) found that search engines are one of key venues for news discovery. Similar observations were generated by (Ulloa & Kacperski, 2022) who also observed the frequent use of search engines for engaging with the news. Other studies highlight the importance of search engines for engaging with breaking news at the time of crisis, such as the COVID-19 pandemic (e.g. Nielsen et al. 2020, 2021). Arguably, the contested elections can be viewed as such a crisis, where search engines are relied on as an important source of news updates.

At the same time, search engines are used not only to access the news, but also other types of content, including content dealing with conspiracies. Based on Google Trends data, Rovetta and Others (2021) demonstrated the intense use of Google for search for conspiracy-related information in the context of COVID. Similarly, PyrhoNen and Bauvois (2020) showed that search engines, such as Google, are often utilized for search for politics-related conspiracies (e.g., pizzagate). Recent audits (eg Makhortykh et al., 2020; Urman et al., 2022) demonstrated that different search engines put various amounts of effort into filtering out conspiratorial information, with some of them – most notably, 'privacy sensitive' DuckDuckGo – being lauded by adherents of conspiracy theories as the ones offering higher visibility to alternative views.

## 2.3   Search Engines and Conspiracies in the News

We suggest that variation in the search results is influenced by three main factors. The first is the search algorithm itself and content moderation practices: different search engines follow different algorithmic logic to search the web and present users with results. Differences in the logic lead to distinct outputs, as consistently demonstrated by relevant comparative research (Makhortykh et al., 2020; Urman et al., 2020). The second factor is the location of the search. It determines what is treated by the search engine as the most relevant results based on IP addresses of search users. This very basic level of personalization, however, can already result in bias: for instance, searches from an American state that is strongly

Democrat-leaning could, for example, result in more liberal news media being promoted. Such location-based differences in the retrieved content were observed by Kliman-silver et al. (2015). A third factor is the actual search term used. Besides the obvious point that search terms are the starting point to request any information, studies have shown that small variations in terms used can lead to significantly different results. This is important considering evidence showing that partisans use distinct search queries for similar issues (Van Hoof et al., 2022). Furthermore recent work by Tripodi (2022) has argued that right-wing elites prime their audiences with 'ideological dialects' – keywords that individuals can use in search engines that result in information supportive of right-wing ideas.

These three factors can lead to differences along three main axes in news results for the Big Lie. The first is the source of the news stories retrieved by the search engines – because different news sources cover stories differently, systematic differences in the sources of present a problem in building a common understanding of the conspiracy. Major discrepancies in the selection of sources by different engines or in different locations on topics such as elections can result in information inequalities and amplify existing ideological divisions. At the same time, previous studies have shown that different search engines prioritize different information sources – we, therefore, hypothesize that (**H1**) *Different search engines present conspiracy news from different sources.* However, little is known about the effect that location can have on the information sources used. We, therefore, ask (**RQ1**) *Do differences in locations lead to dissimilar sources?*

The second axis is the relevance of the result: search engine algorithm, location of the search, and search query used can all impact the degree to which outputs are relevant to a specific subject. The relevance of search results reflects how well the engine fits the expectations of its users (i.e., finding the information that the users want to find). However, in the case of ontologically contested subjects (e.g., conspiracies), the engine also has to deal with the multitude of opinions on the subject (e.g., whether to give visibility to materials promoting/debunking conspiracies or avoiding discussing the conspiracy). The choices made by the search algorithms in this particular context are essential for defining their role in the public sphere and the way their might promote or undermine different types of democracies (e.g. liberal or participatory; Helberger, 2019). For this output, we specifically focus on differences between search engines, asking (**RQ2**) *Do differences in search engines lead to differences in result relevance when searching for information on the Big Lie conspiracy?*

Third is the level of debunking present in news articles – i.e., the main focus of the study. Whether search engines prioritize stories that support, fact-check, or debunk the Big Lie has dire implications for the role that these intermediaries play in the spread of the conspiracy. Urman et al. (2022) show, for example, the tendency of DuckDuckGo to promote historical conspiracies (e.g., the New World Order theory). We, therefore, hypothesize that (**H2**) *DuckDuckGo will present significantly more news pieces supporting or not fact-checking Big Lie claims than other search engines.* Few studies have investigated how location influences the degree of debunking present in conspiracy-covering news articles – we, therefore, ask, (**RQ3**) *Do differences in location lead to differences in news article conspiracy debunking when searching for information on the Big Lie conspiracy?* Lastly, we address the role that specific search queries have in terms of conspiracy debunking. As Van Hoof et al. (2022) have shown, partisans use distinct keywords to search for specific topics. Furthermore, on the output level, Tripodi (2022) argues that far-right politicians and media elites prime their publics with specific events and keywords that reaffirm specific narratives. As part of this conservative information sphere, the above-mentioned actors prime their followers with 'ideological dialects' (Tripodi, 2022) – i.e., 'insider speech' keywords that, when used in search engines, will return information that supports an ideologically-driven view of the world, including conspiracy theories. We therefore hypothesize that (**H3**) *'Insider keywords' will result in more conspiracy-supporting news articles.*

## 3    Data and Method

### 3.1    Data Collection

To collect data for the study, we conducted a virtual agent-based algorithm audit of three search engines: Google, DuckduckGo, and Bing. The choice of these search engines was based on them holding the biggest share of the search engine market in the US. We decided not to include other relatively popular engines such as Yahoo and T-Online, as they often employ the same underlying algorithm as the initial three.

   Unlike other approaches for search engine audits (for an overview see Bandy, 2021), virtual agent-based audits rely on simulating human activity for generating inputs for the search engine. Such a simulation is implemented using software that imitates user actions (e.g., entering a search query) and then records the outputs (for further details on the approach, see Ulloa, Makhortykh, & Urman, 2022). This approach facilitates controlling for the effects of search personalisation (i.e. search algorithms customizing their outputs for individual users based on, for instance, their location and timing Hannak et al., 2013) and search randomisation (i.e. algorithms randomly reshuffling the search outputs, for instance, as part of A/B testing Urman et al., 2021).

To conduct the audits, we used a self-designed cloud infrastructure deployed via Amazon Elastic Compute Cloud (EC2) and made of CentosOS virtual machines. The machines were located in the three regions of the EC2 - Ohio, Northern California and London - so the agents would have IP addresses similar to humans from these areas. The choice of the regions was attributed to our interest in comparing search outputs for the regions with different political leanings (i.e., California as Democratic and Ohio as more conservative) with the UK (London) being used as a control region. While there are arguably states that are more pro-Republican than Ohio, our locations were limited by the choice of the regions provided by Amazon.

Two browsers (i.e., Firefox and Chrome) were installed on each machine and in each browser (i.e., the agent), two browser extensions were installed. The first extension - the bot - emulated browsing behaviour by opening the news search page of one of four search engines (in all cases the .com version was used for consistency), entering the query, and then cleaning data accessible by the browser and the engine's JavaScript to prevent earlier searches affecting the subsequent ones. The second extension the tracker - collected the HTML from each page visited by the bot.

To collect news articles regarding the Big Lie, a series of queries were chosen to reflect both a mix of 'ideological dialects' (Tripodi, 2022) regarding voter fraud, as well as more general and less ideologically motivated discussions on voter fraud allegations. Table 1 lists all the search terms used, the specific element of the conspiracy theory it is referring to, as well as their categorization as either 'Insider search term/Ideological dialect' or 'Broad search term'/ We used the insider-speech nickname of the voter fraud theory with the term 'stopthesteal', as well specific instances of alleged fraud, such as 'sharpie marker ballot arizona', 'water leak fulton county', 'poll watchers blocked', 'overnight vote change', 'signature match mail in voting', 'dominion voting systems change votes', 'suitcases votes found', 'absolute proof movie'. As a comparison, we also included the general queries, such as 'voter fraud 2020' and 'dead votes 2020 elections' – i.e., terms that could be used to find information on voter fraud for any election simply by changing the year.

Data collection was conducted on the 13th of March 2021. We simultaneously deployed 360 virtual agents equally distributed between search engines and locations, meaning there was a total of 120 agents within a given location and 40 agents per search engine.

## 3.2   Data Analysis

For each query, we extracted the top twenty results that reappeared throughout the 12 engine-location combinations (e.g., Ohio-Google or Bing-UK). By focusing on the top twenty results that constantly reappear, we are able to reduce the noise that is produced by randomization within search engines (Ulloa et al., 2022; Urman et al., 2021) and focus on the results that the users are most likely to be exposed to. The content within these URLs was then manually coded by a set of three trained coders.

Table 1: Search Term Explanation and Categorization

| Query | Conspiracy Theory Explanation | Categorization |
|---|---|---|
| stopthesteal | Broad slogan adopted by Trump supporters and individuals believing the election was fraudulent. This slogan was used as a rallying cry, as well as an organizational hashtag. | Insider search term/ Ideological dialect |
| sharpie marker ballot arizona | Refers to the theory that Republicans were handed sharpie markers to label their ballot. The theory holds that ballots filled out with sharpies were then not counted due to technical errors. The particular incident took place in Maricopa County, Arizona, where a lawsuit followed suit, as well a #Sharpiegate hashtag. Arizona, and specifically Maricopa County, were crucial for Joe Biden's victory. | Insider search term/ Ideological dialect |
| water leak fulton county | Refers to the theory that pro-Biden votes were 'dumped' in a vote-counting centre in Fulton County, Georgia, after voting officials were evacuated due to a burst water pipe. The theory holds that this evacuation eliminated Republican oversight in the polling station, allowing democrats to introduce ballots in favour of Joe Biden. Georgia, and specifically Fulton County, were crucial for Joe Biden's victory. | Insider search term/ Ideological dialect |
| signature match mail in voting | Refers broadly to the conspiracy that Democrats made use of insecure mail-in voting to commit fraud. One of the alleged way to commit such fraud was by a lack of signature matching systems to verify the veracity of the person casting the vote. | Insider search term/ Ideological dialect |
| poll watchers blocked | Refers to the theory that Republican poll watchers were blocked from monitoring the vote-counting process in polling stations across Pennsylvania and Michigan, key states for Joe Biden's victory. The theory holds that such a lack of oversight widespread fraud. | Insider search term/ Ideological dialect |
| overnight vote change | Refers to the theory that there was an organized switching of votes that occurred overnight in Wisconsin, Pennsylvania, and Michigan. The theory developed as a product of the fact that Joe Biden overtook Donald Trump in these states during the night, as mail-in votes were counted and released in large batches. | Insider search term/ Ideological dialect |
| dominion voting systems change votes | Refers to the theory that Dominion voting systems systematically deleted votes for Donald Trump, or switched them to Joe Biden. Many variants of this theory exist - some hold that these discrepancies happened because the machines were connected to the internet, others that internal rounding errors led to a manufactured Biden advantage. Such unfounded claims led to numerous lawsuits filed against Dominion by prominent Trump supporters. | Insider search term/ Ideological dialect |
| suitcases votes found | Refers to the conspiracy that in Fulton County, Georgia black suitcases of votes were secretly introduced to a voting station after poll watchers had left. | Insider search term/ Ideological dialect |
| absolute proof movie | Refers to the Absolute Proof 'documentary' that Trump ally and prominent voter fraud spokesman Michael Lindel produced. The 'documentary' features details of numerous conspiracy theories of how the election was fraudulent and ultimately stolen from Donald Trump. | Insider search term/ Ideological dialect |
| voter fraud 2020 | Refers to the general idea of voter fraud taking place in the American 2020 election. It is purposefully vague and ideologically neutral. | Broad search term |
| dead votes 2020 elections | Refers to the idea that the registrations of dead citizens were used to fraudulently cast additional ballots. This is usually among the first claims made when an election is contested as fraudulent, and not specific to the 2020 American election. | Broad search term |

Three main variables were coded:

**Website source:**

- Results were coded by whether they were news media – broadly defined as content informing on current events – or whether they were not news media (for example, Wikipedia pages or government pages). Non-news media results were excluded.

**Degree of focus on the conspiracy:**

- Article focuses *almost exclusively* on the conspiracy.

- Article *mentions* the conspiracy, but the conspiracy is not the focus.

- Unrelated to the conspiracy — article shows up because of some mismatch of the word in the search engine, but is actually about something completely different.

- Avoids mention of the conspiracy – article focuses on a topic that is close to the election conspiracy (e.g. the passing of Republican voter integrity bills), but makes no mention of claims of fraud.

**Degree of conspiracy debunking:**

- 'Debunking': article (mostly) dedicated to providing an account of why the conspiracy/element of conspiracy is incorrect. This also may include instances of alternative explanations being provided.

- 'Fact check': When discussing the conspiracy, the authors state that there is no evidence, or that external sources have confirmed that this is not true, negating the conspiracy. The article, however, does not go into depth as to why this is not true and does not provide an alternative explanation for the conspiracy.

- 'Mentions conspiracy': The article mentions conspiracy but does not clearly support or deny the conspiracy.

- 'Promotes conspiracy': The article actively either provides arguments for the conspiracy or makes an explicit assumption that the conspiracy is correct.

Figure 1: Overlap in the Domains Presented by Each Search Engine per Location

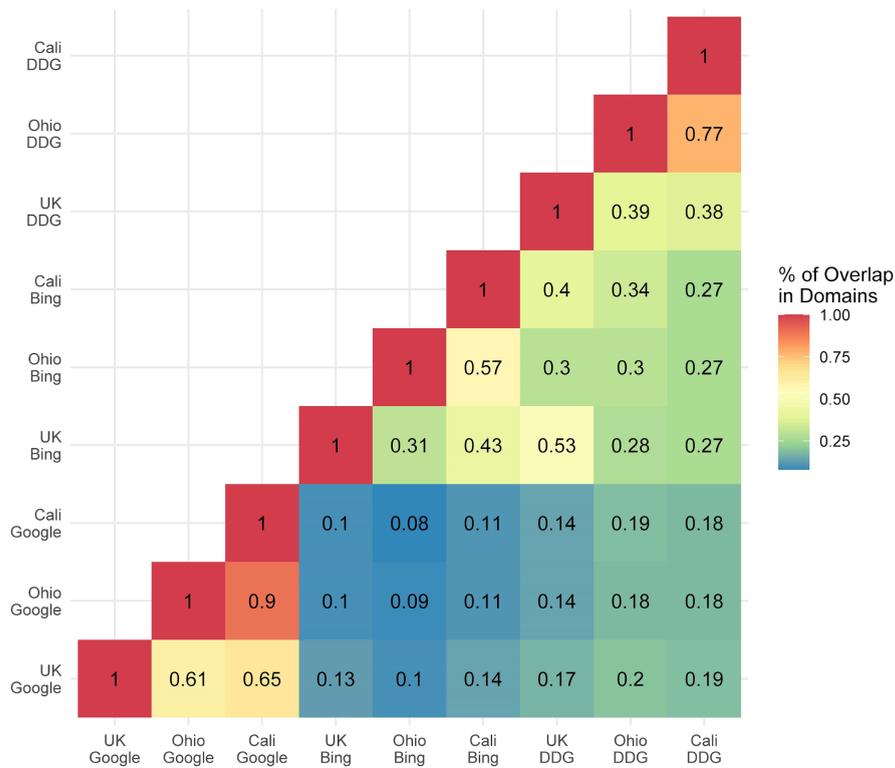

## 4    Results

We begin the discussion of the results by examining whether different search engines present users with similar sources (**H1**), and whether this varies by location (**RQ1**). Figure 1 shows a heat map representing the overlap in domain results by each engine (Google, Bing, DuckDuckGo) in three different locations (UK, Ohio, and California), with a percent overlap presented in each tile. Here, it is important to note that each number represents overlap in domains, and not unique URLs, thus presenting a conservative picture of differences between results (different articles from the same source count as an overlap). Figure 1 shows that most similarity occurs within search engines. The clearest example is that of Google: conspiracy searches from California and Ohio shared 90% of the same domains. A similar picture is painted with DuckDuckGo (albeit with a large drop), with 77% overlap in domains for California and Ohio. This means that two individuals using Google or DuckDuckGo in these two US states can expect to find information from similar sources.

Nevertheless, differences in results begin to surge as we compare with the United Kingdom, and when we look at Bing. For Google, only 65% of domains were the same for California and the UK, and 61% for Ohio and the UK. Within DuckDuckGo, overlap is further reduced, with the percentage of shared domains dropping below 40% between the UK and the two American states. While Bing confirms the pattern previously identified – the highest domain overlap between Ohio and California (57%) – this overlap is significantly lower. Domains shared between the US states and the UK also drop to 31% and 43%.

Figure 1 shows the overlap between search engines within and across different locations. Here, numbers drop significantly: results from Google in the UK have only 13% and 20% overlap to the domains from Bing and DuckDuckGo despite the same location, a picture that is repeated when searching within Ohio (10% and 18% overlap) and California (10% and 14% overlap). The share of domains is higher for Bing and DuckDuckGo within the UK, with 53% overlap; however, the remaining within-location across-engine overlap stays below 35%.

These results suggest that differences in search engines' performance are the driving factor in the variation in prioritized sources, confirming **H1**: most the domains produced by Google, DuckDuckGo, and Bing were completely different, despite searches being identical, and conducted within the same location. This, however, does not mean that the search location does not matter (**RQ1**) – within search engines, changes in the location lead to a significant source variation. In particular, we see that these differences are lower when searching within the US (Ohio and California), and exacerbated when compared to the UK.

Figure 2: Distribution in (A) Conspiracy Focus and (B) Conspiracy Debunking by Search Engine

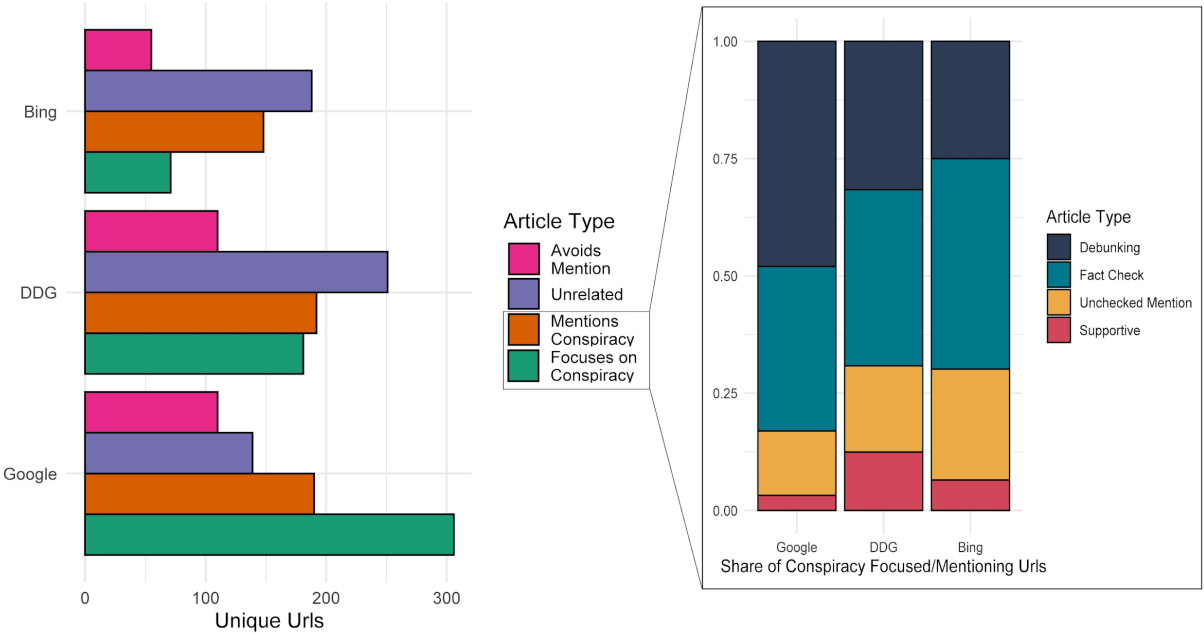

Table 2: Detailed Output of Distribution in (A) Conspiracy Focus and (B) Conspiracy Debunking by Search Engine

Degree of Conspiracy Focus

| Engine | Focuses on Conspiracy | Mentions Conspiracy | Unrelated | Avoids Mention | Total |
|--------|----------------------|--------------------|-----------|---------------|-------|
| Bing   | 71                   | 148                | 188       | 55            | 462   |
|        | 15,37%               | 32,03%             | 40,69%    | 11,90%        | 100%  |
| DDG    | 181                  | 192                | 251       | 110           | 734   |
|        | 24,66%               | 26,16%             | 34,20%    | 14,99%        | 100%  |
| Google | 306                  | 190                | 139       | 110           | 745   |
|        | 41,07%               | 25,50%             | 18,66%    | 14,77%        | 100%  |

Degree of Conspiracy Debunking

| Engine | Debunking | Fact Check | Unchecked Mention | Supportive | Total |
|--------|-----------|-----------|-------------------|-----------|-------|
| Bing   | 54        | 97        | 51                | 14        | 216   |
|        | 25%       | 44,91%    | 23,61%            | 6,48%     | 100%  |
| DDG    | 117       | 139       | 68                | 46        | 370   |
|        | 31,62%    | 37,57%    | 18,38%            | 12,43%    | 100%  |
| Google | 238       | 174       | 68                | 16        | 496   |
|        | 47,98%    | 35,08%    | 13,71%            | 3,23%     | 100%  |

Figure 2 addresses **RQ2** on the relative focus news articles appearing as a result of a Big Lie information search had on the conspiracy, and **H2**, on how they treated the veracity of the allegations (precise numbers can be found in Table 2). Panel A of Figure 2 shows how many of the news articles in the results 1) focused explicitly on the election fraud conspiracy, 2) mentioned the conspiracy as part of a bigger/unrelated story, 3) were completely unrelated to the conspiracy, and 4) were plausibly related to the conspiracy, but did not mention it. The results show significant differences between search engines: a large portion of Google's results (41%) focused on the conspiracy, a number much lower for both DuckDuckGo (24.5%) and Bing (15.3%). The distribution for conspiracy mentions were similar across the three search engines, ranging from 32% (Bing) to 25.5% (Google). With unrelated articles we again observe large differences - Bing leads this category with 40.6% of its results, followed by DuckDuckGo 34%, and a much lower number from Google (18.7%). The 'avoids mention' category is the smallest for all, representing between 11.9% (Bing) and 14.9% (DuckDuckGo) of unique results.

Panel B of Figure 2 focuses on articles that either mention or focus on the Big Lie and examines whether these articles 1) completely debunk the conspiracy, 2) briefly fact-check it, 3) mentions the conspiracy without adding whether it is actually true or not, or 4) support the conspiracy. While factchecking results are represented relatively equal amongst search engines (i.e., between 35 and 45%), large differences emerge in the degree of debunking. Almost 50% of relevant Google results focus on the thorough debunking, a number far lower for DuckDuckGo (32%) and Bing (25%). Similar are differences in the number of results supporting the conspiracy: for Google, this number is 3%, which doubles for Bing (6%), and quadruples for DuckDuckGo (12%). This confirms H2 that suggested that DuckDuckGo would present much higher levels of conspiracy-promoting results.

Figure 3: Conspiracy Coverage by Articles Focused on the Conspiracy

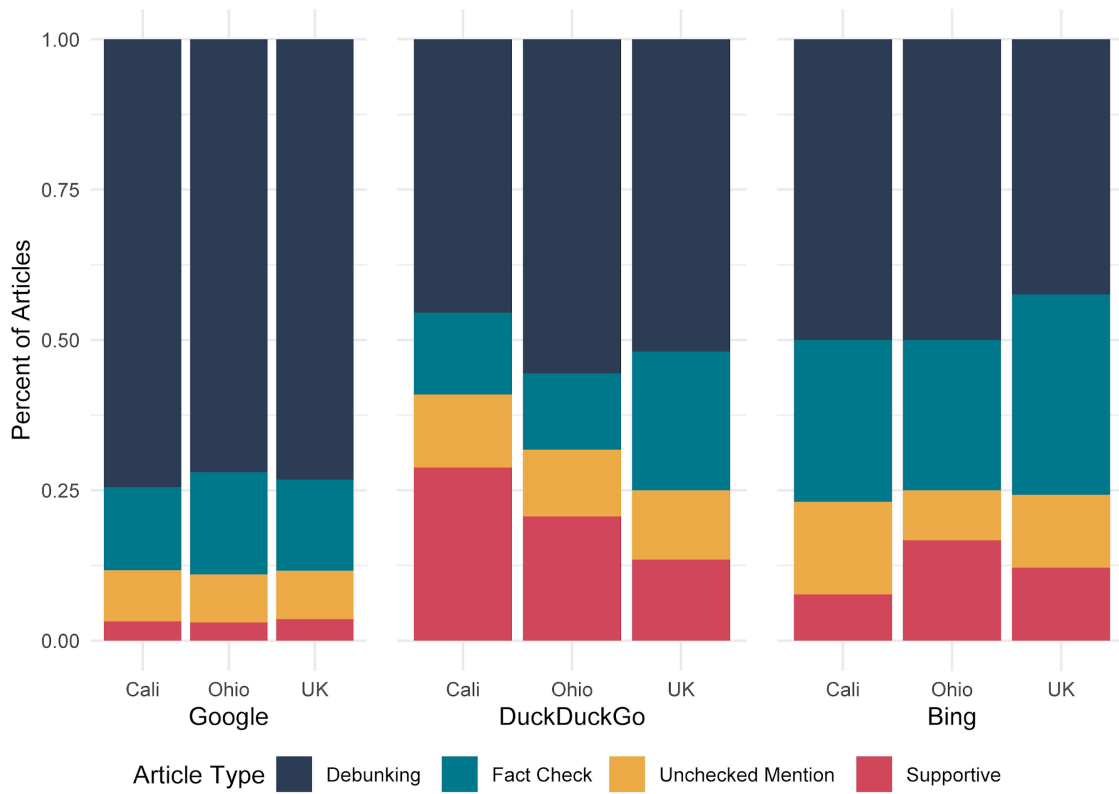

Figure 3 offers a closer look at the news stories that are exclusively focused on the conspiracy, arguably the most important type of search results when it comes to the propagation of the Big Lie.

It divides search engines by location, shedding light on whether the differences identified in Figure 1 translate into differences in the proportion of debunking results, and allowing us to answer **RQ3** on the relationship between location and conspiracy debunking. We observe that when looking exclusively at news articles covering the Big Lie, the share of debunking results rises: for Google, they account for about 75%, and for DuckDuckGo and Bing around 50%, doubling from the results in Figure 2. This is teamed with a decrease in the number of articles that 'only' offer a fact check or simple negation. Nevertheless, we also see a rise in articles supporting the conspiracy, a change most evident for DuckDuckGo and Bing, where this content averages over 20% of unique articles for DuckDuckGo (up from 12.43%), and 12% for Bing (up from 6.48%).

Because Figure 1 showed differences in domains by location, it is plausible to expect differences in conspiracy coverage between them. DuckDuckGo confirms that this is indeed the case – the percent of articles supporting the conspiracy or not fact-checking it at all varies significantly between California, Ohio, and the UK. For California, such articles account for almost 40% of the results, while they fall to 30% in Ohio (which also received a greater share of debunking stories), and 25% in the UK. For Bing, differences between locations also exist: searches in Ohio resulted in more than twice as many articles supporting the conspiracy than in California. For Google, however, we see a stable distribution of stories, echoing the relevant overlap displayed in Figure 1.

Figure 4: Search Results by Query and Engine

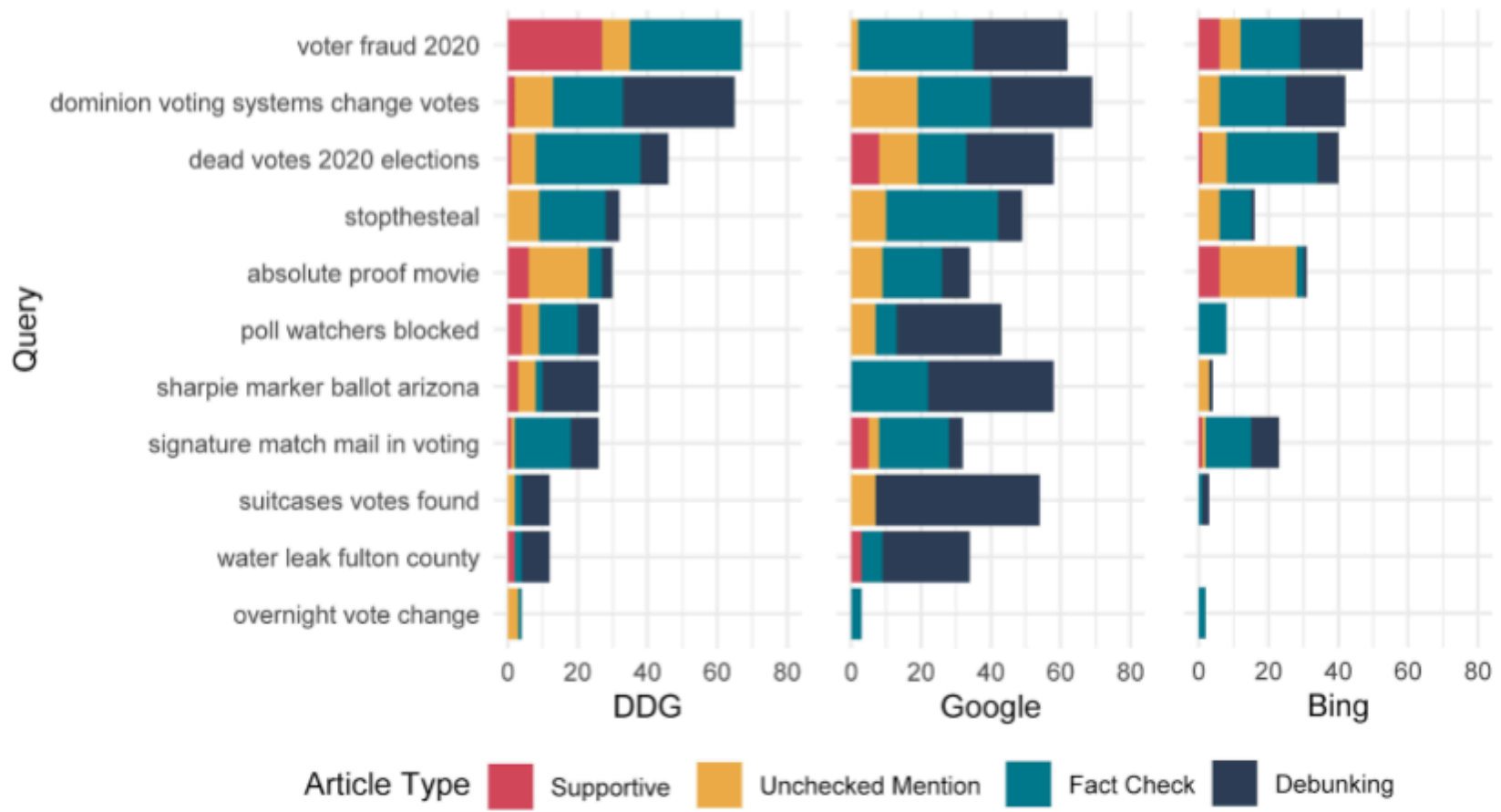

Lastly, to answer **H2** on the effect queries can have on the Big Lie search results, Figure 4 plots the distribution of conspiracy-related results by queries for each search engine. In the figure, we distinguish between words that were added as a form of 'ideological dialects', as we hypothesized that these would result in higher conspiratorial information, and two 'control' searches ('voter fraud 2020' and 'dead votes 2020 elections'). For both DuckDuckGo and Bing, we see that the number of articles explicitly supporting the Big Lie are highest for the 'voter fraud 2020' and 'absolute proof movie'. For Google, the pattern is distinct, with no supportive articles for these queries; instead, conspiracy results appear as a result of 'dead votes 2020 elections' and 'signature match mail-in voting'.

**H2** suggested that narrower, more niche 'ideological dialects' as search terms would lead to results supporting the election fraud conspiracy theory. However, we find little evidence that ideological dialects are related to more conspiratorial outputs in search results. In fact, the results show much of the opposite
– the queries with the highest number of articles supporting the conspiracy are by far the 'control' terms: 'voter fraud 2020' (for DuckDuckGo and Bing), and 'dead votes 2020 elections' (for Google). 'Stop the steal', which has become the rallying cry for the movement that believes Biden won through fraud, resulted in zero results supporting the Big Lie.

While our results do not provide strong evidence that ideological dialects lead to conspiracy supporting information, our findings do not offer a clear alternative explanation, either. For Bing, but especially DuckDuckGo, the broad 'voter fraud 2020' query results in the largest number of conspiracy supporting outputs. However, for Google, the same query brings about almost exclusively debunking and fact-checking outputs. A similarly broad query – 'dead votes 2020 elections' – results in the highest number of conspiracy-supporting information for Google, while among the lowest for DuckDuckGo and Bing, so we cannot claim that ideology-agnostic queries always lead to more conspiratorial outputs.

Lastly, we do see that Google does a significantly better job at debunking concrete tropes associated with the Big Lie. The more niche queries, such as 'water leak fulton county', 'suitcases votes found', 'sharpie markker ballot arizona', and 'poll watchers blocked', which correspond to very specific events within the conspiracy, are met with a large portion of debunking news articles. The situation, however, is rather different for the two other search engines.

## 5 Discussion

This study has investigated how search engines served as a gateway to news on the conspiracy theory that the US 2020 presidential election was won through voter fraud. It analyzed how the news retrieved by search engines engaged with conspiratorial claims. Using an audit methodology that employed virtual agents to systematically query search engines, the study compared results from three different locations (California, Ohio, and the United Kingdom), across three different search engines (Google, DuckDuckGo, and Bing), and for a variety of search queries.

As a starting point, our study investigates whether different search engines and locations lead to a variety in prioritized news sources. We find that the largest source of variation comes from using different search engines. Despite using the exact same search terms at the exact same time, the sources presented by Google were different to those presented by Bing and DuckDuckGo – at most, the engines shared 20% of news sources. These results reiterate earlier concerns about the impact of search engines on information access, in particular the anxieties about them amplifying information inequalities among citizens. Furthermore, we find that *within* search engines, results exhibit different patterns depending on the location. For Google, sources for California and Ohio (both within the US), overlap in 90% of cases – a number that drops to 77% for DuckDuckGo and 57% for Bing. These numbers are lower when we analyze source overlap between these the US and the UK, with only around 63% overlap for Google, 39% for DuckDuckGo, and 40% for Bing. It confirms the intuitive assumption that source overlap should be higher within the US locations than between the US and the UK, as search engines likely prioritize different national media. Nevertheless, these numbers are still worryingly low, as they suggest that 23% and 43% of news sources are different for DuckDuckGo and Bing, despite using the same queries, at the same time, and within the same country. Such differences could lead to distinct understandings of the events surrounding the Big Lie conspiracy and potentially increase polarization in the already divided society.

Turning to the actual content of search results, we find a large gap in the relevance of the information for the Big Lie between search engines. Only 41% of results for Google focus explicitly on the conspiracy, a number drops to 24% for DuckDuckGo, and 15% for Bing. In contrast, the number of news articles that are completely unrelated to the conspiracy stands as high as 40% for Bing, followed by 34% for DuckDuckGo and 18% for Google. This shows differences in the treatment of epistemically challenged claims by individual engines. We also show that all three search engines retrieved articles that were related to the conspiracy, but avoided mentioning it, with this category containing around 14% of results.

The news media plays a key role in both spreading beliefs in conspiracies and debunking them. This study sought to uncover how different search engines promoted articles that either debunked, fact-checked, supported, or simply mentioned the conspiracy. The results show that for all search engines, fact-checking is the most popular form of engagement with the Big Lie. While such an approach is arguably better than promoting the conspiracy, studies have shown that simple fact-checks can lead to the spreading of the conspiracy (Cook et al., 2015; Druckman & Mcgrath, 2019; Walter & Murphy, 2018). From this literature, it is arguable that simple negations, the most common form found in search engines, would not be enough to stop the spreading of conspiracies. The large presence of simple negations is therefore a worrying takeaway, considering the role search engines play as intermediaries to the news. Even more worrisome, however, are the differences in debunking: while almost 50% of conspiracy-related outputs for Google provide detailed debunking of the conspiracy, this number drops to 30% for DuckDuckGo and 25% for Bing. These numbers are worrying considering that for DuckDuckGo and Bing, articles not fact-checking the conspiracy or actively promoting it are more numerous than debunking results.

To better understand the role that search engines play in the dissemination of conspiracy information, we examined how conspiracy-focused news treats information about conspiracies. It is likely that these articles, rather than ones that simply mention the conspiracy, would provide the information capable of changing individual opinions Here, the picture improved for the largest search engine – Google: 75% of articles focusing on the conspiracy debunked the Big Lie, with only 3% supporting the conspiracy. For DuckDuckGo and Bing, on the other hand, the rise in debunking articles (from 25% to 50%) is accompanied by a sharp increase in the number of articles openly supporting the conspiracy (25% for DuckDuckGo and 15% for Bing). These results suggest that articles reiterating the Big Lie claims are far from absent in search engine results, with two out of three major search engines consistently prioritising them.

Such a relatively high rate of articles openly supporting the voter fraud conspiracy on DuckDuckGo is especially worrisome considering the profile of the public using it. Promising to be a privacy-sensitive alternative to Google, this engine has drawn individuals that are arguably the least trusting of the political system (Urman et al., 2022) – and, consequentially, most likely to believe that the election was stolen. Such acute differences also foreground a fundamental tension between quality and monopoly. While this study shows that Google outperforms Bing and DuckDuckGo when it comes to providing users with sound news on conspiracy theories, the company has been criticised for its monopolistic hold over the search engine market. A more profound conversation should take place on the trade-offs between search engines providing users with misinformation and giving a single platform even more power.

Lastly, we examined the link between search queries and engagement with conspiracy theories. Past research has argued that search queries form an integral part of finding political information – not only do partisans use different queries when searching for similar information (Van Hoof et al., 2022), but ideologically-driven queries might result in different results (Tripodi, 2022). In our audit, we included a mix of broad conspiracy search queries, along with more niche ideologically-driven queries. Nevertheless, our results show that more ideological queries were not actually linked to a higher proportion of conspiracy-supportive results. In fact, broader queries routinely received more results supporting the election fraud conspiracy theory, likely a product of a larger pool of articles available. These findings align with earlier studies calling for caution against extending claims about filter bubbles to search engines (Krafft, Gamer, & Zweig, 2018).

This study is not without its limitations. First, the timing must be recognized: while we believe that the Big Lie (and its political consequences) was still a major news item at the time, the audit took place in mid-March 2021, several months after the 2020 election. Therefore our results are not a real-time representation of the news coverage of the conspiracy, but rather a retrospective look. Second, while we took significant steps to control for the effects of randomization and personalization, it also implies that we used 'empty' accounts to mimic more complex reality. In the 'real world' environment, search engines make use of extensive data on users to personalize search outputs for individuals. Lastly, while we strove to provide an ideology-agnostic set of queries, it could be reasonably argued that 'anti-conspiracy' queries should have also been included. Such inclusion would have allowed for a more nuanced comparison in results based on the ideological proclivity of keywords.